# Space-filling discrete helices

**Jayanth R. Banavar[1], Achille Giacometti[2,3], Trinh X. Hoang[4], Amos Maritan[5], Tatjana Škrbić[2, *]**

[1] University of Oregon, Department of Physics and Institute for Fundamental Science, Eugene, Oregon, USA

[2] Ca' Foscari University of Venice, Department of Molecular Sciences and Nanosystems, Venice, Italy

[3] European Centre for Living Technology (ECLT), Ca' Bottacin, Dorsoduro 3911, Calle Crosera, 30123 Venice, Italy

[4] Vietnam Academy of Science and Technology, Institute of Physics, Hanoi, Vietnam

[5] University of Padua, Department of Physics and Astronomy, Padua, Italy

[*] Corresponding author email: tatjana.skrbic@unive.it

## Abstract

**Proteins are linear chain molecules that play a central role in life and health. Protein native state folds are modular assemblies of space-filling building blocks of α-helices, β-sheets and tight turns. Here we deduce the structures of a countable set of space-filling helical forms of a uniform discrete thick string from first principles with no additional input or adjustable parameters. These forms occur in correspondence with the natural numbers, loosely analogous to the energy levels in a Bohr atom. We find the remarkable result that one of these helical forms is an excellent candidate for an α-helix through seemingly improbable quantum chemistry coincidences that fit the geometrical requirements. Our work suggests that geometry and chemistry are complementary ways of looking at proteins and suggests a route for developing a unified framework for understanding proteins.**

*Keywords:* poking, protrusion, symmetry, hydrophobicity

Linus Pauling, Robert Corey, and Herman Branson [1] deduced the molecular structure of the α-helix, a key building block of proteins using quantum chemistry and recognizing the pivotal role of hydrogen bonds. A straightforward extension of their techniques to the level of a full protein is daunting because of the very large number of distinct putative sequences each of which must be studied individually, and the sheer scale of the complexity - thousands of atoms and myriad additional interactions. Recently, amazing strides in AI [2-5] have solved the sequence-structure problem, yet a simple understanding has remained elusive. Protein native state structures are modular and are made of secondary motifs of α-helices and β-sheets. These building blocks and the interior of folded protein structures expel water from within and tend to be well-packed with protein atoms. Space-filling within the α-helix is believed to occur by maximizing the self-interaction of backbone atoms with support from hydrogen bonds [6].

**The genius of Linus Pauling**

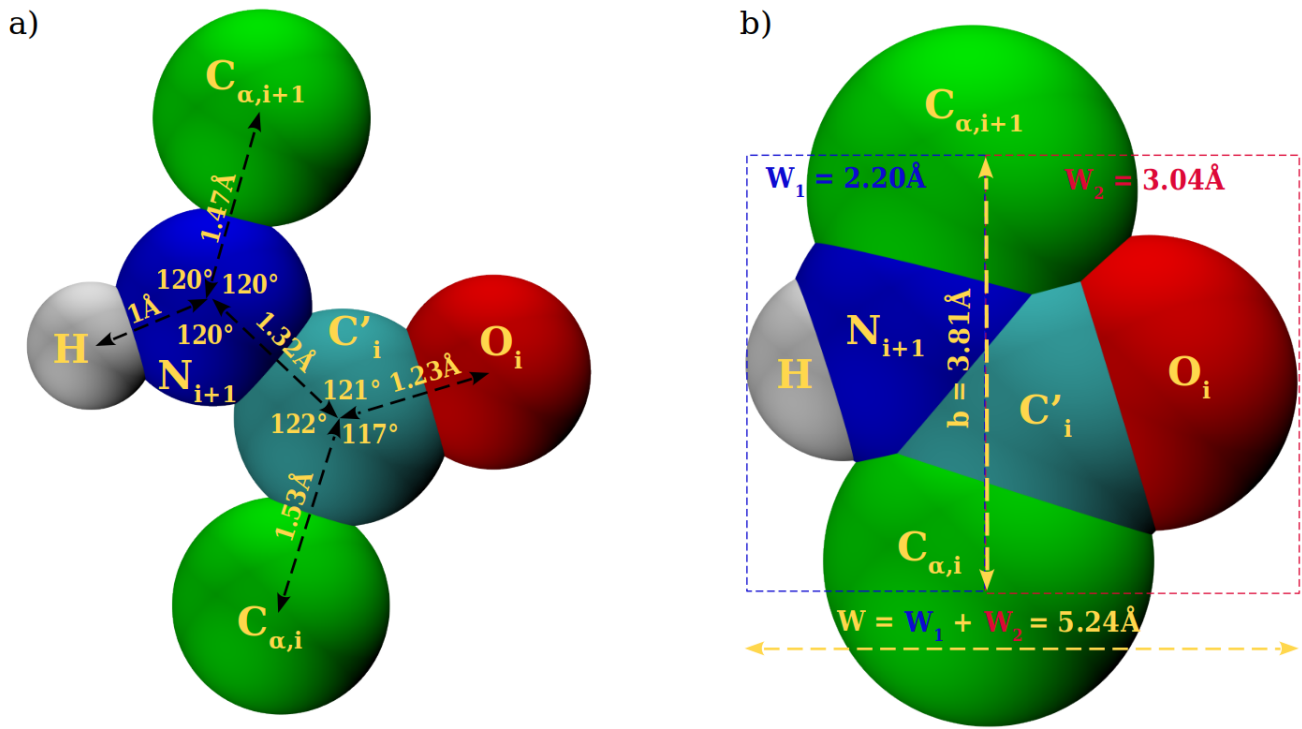

Figure 1: Sketch of the centers of atoms lying in a planar peptide bond (adapted from Pauling [1]). In (a), the sizes of the atoms are half of their van der Waals diameter for clarity and in (b), the actual sizes are shown. The distance and angle constraints are those deduced by Pauling using chemical constraints and X-ray data on small peptides. $C_\alpha$ atoms are shown in green, carbon atoms other that $C_\alpha$ are turquoise, oxygen atoms are red, nitrogen atoms are dark blue, while hydrogen atoms are shown in light gray color.

Pauling and his colleagues [1] proposed that a characteristic attribute of proteins is repeating secondary structure elements like the α-helix arising from the self-interactions of the common backbone of all proteins. They determined the α-helix structure by integrating chemical constraints and X-ray data on small peptides to infer distances and angles between atoms in proteins, along with model building, to propose a helix stabilized by internal hydrogen bonds. The carbonyl oxygen (C=O) of one amino acid peptide plane forms a hydrogen bond with the amide hydrogen (N-H) of an amino acid peptide plane placed four residues ahead in the chain. Interestingly, Pauling found two acceptable models of a helix, one of which has 3.7 amino acids per turn and a pitch, P, of around 5.4 Å (denoted as the α-helix in Table 1) in excellent accord with subsequent structural studies.

The molecular structure of a protein is governed by the planarity and rigidity of the peptide bond arising from its partial double-bond character. In essence, the peptide bond is a rigid plate, with fixed relative locations of the associated backbone atoms. The quantum chemistry underlying the planar peptide bond [1] ensures that the $C_\alpha$ atom is an excellent proxy for the equidistant points along the string (Figure 1). The pseudo-bond length b or the distance between successive $C_\alpha$ atoms is approximately constant and equal to 3.81Å [7], remarkably independent of *most* local rotational conformations characterized by Ramachandran angles [8]. Treating the atoms as hard spheres with their individual van der Waals radii, one finds that that the planar peptide bond forms a space-filling plane with a lateral spatial extent b and transverse width W (Figure 1b).

While Pauling's work laid the foundation of molecular biology and is an amazing solution to the structure of the α-helix, the quantum chemistry approach is not tractable when one must deal with the size and complexity of a full protein along with innumerable distinct sequences.

**AI comes to the rescue**

Christian Anfinsen [9] established that upon repeated denaturation and renaturation, a protein reliably returns to the same folded state, demonstrating a unique mapping between its amino acid sequence and its biologically functional structure. This foundational result poses a central question in molecular biology: how can one predict the native conformation of a protein solely from its primary sequence? Recent advances in artificial intelligence (AI) [2-5] have made significant progress in addressing this sequence-to-structure prediction problem. These developments leverage extensive structural databases, multiple sequence alignments, and powerful machine learning methodologies—including novel deep neural network architectures. AI models are essentially black boxes: they learn correlations but don't offer mechanistic explanations. AI has made structure prediction vastly more accurate and accessible, but it remains a descriptive tool, not yet an explanatory theory. AI has shown us the patterns—but not yet the principles [10].

**Kepler and space-filling**

Optimal arrangements, such as the highest packing fraction of atoms in a crystal, are of relevance in diverse disciplines [6,11-13]. The fcc crystal structure is an optimal packing of hard spheres and is adopted in diverse situations such as the crystalline structure of common salt (NaCl) and by a grocer packing her apples.

In a biological context, the optimal shape of a *continuous* thick string subject to generic compaction, which maximizes the radius of a tube enveloping it, was found to be a space-filling helix (Figure 2) [14]. The analysis in [14] invoked the elegant concept of local and non-local three-body radii and yielded a dimensionless pitch to radius ratio $\eta=P/(2\pi R) = 0.3998\ldots$ for the continuum space-filling helix, where P is the helix pitch and R is the radius of the helix. This geometry is in very good accord with that of the protein α-helix.

The translationally invariant helix is the optimal geometry because local space-filling requires a curve for the tube axis with a *constant radius of curvature* equal to the thickness, all along the string to ensure uniform space-filling at the local level. The requirement of non-local space-filling then determines the optimal value of η.

Following Ref. [14], a simple model of the α-helix is a discrete string of equally spaced interaction centers (representing the $C_\alpha$ atoms) curled into a space-filling helix whose envelope has no hole in the center and no space between successive turns with the backbone atoms approximately filling the space within the helix thereby occluding water from any exposed hydrophobic surfaces. Strikingly, the same helical geometry maximizes the entropy of the solvent spheres, when a flexible tube is surrounded by hard spheres of vanishingly small radius [15-18]. In the continuum limit, there is no characteristic length scale thereby restricting the prediction to just one dimensionless quantity, the pitch to radius ratio of the helix.

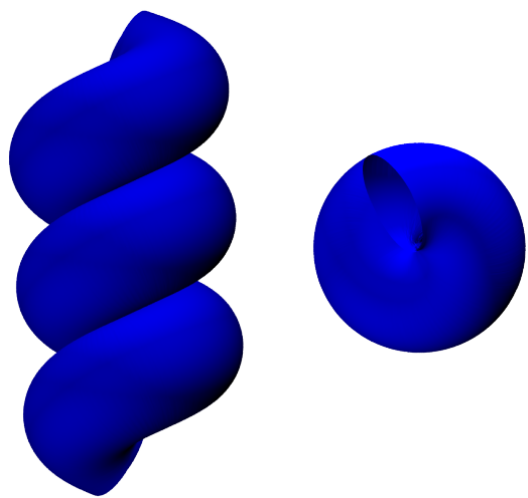

Figure 2: Two views of a space-filling continuum helix. The left part of the figure shows a side view of the non-local space filling between successive turns. The right part of the figure shows the top view of the helix with no empty space in the center because of maximal local space-filling.

**Quantized discrete space-filling helices**

We will unleash the full potential of the continuum space-filling helix by extending it to a *discrete* thick string [19-23], with equally spaced $C_\alpha$ atoms (interaction centers for the protein backbone), labeled 1, 2, 3… i-1, i, i+1……, with a bond length b, lying on a *continuous* translationally invariant helical curve. Here we determine the attributes of a countable set of quantized space-filling discrete helices, each of which is a discrete realization of the continuum helix. We indicate the orientation of a helix by means of its axis (see Figure SI 1 in Supplementary Information). The helical curve denotes the axis of the tube that is curled into a helix. The continuous helical curve and the helix characteristics are fully specified from knowledge of the locations of four successive $C_\alpha$ atoms of a discrete helix.

There are four requirements for the space-filling of a tube in a helical conformation [19-23]. First, one requires local tight coiling of the helix [14] with no empty space in the center of the helix. This arises only when the tube radius $\Delta = R_{curv}$, the radius of curvature of the helix [24]. The coordinates of four successive points of a discrete chain can be used to deduce the radius of curvature. Second, we require that successive turns of the helix track and lie alongside each other with neither an overlap nor empty space between them. This occurs when the relevant circular cross sections of the tube in adjacent turns with centers i and j just touch edge on edge with the distance between their centers exactly equal to 2Δ. This is reminiscent of Kepler's sphere packing constraint, wherein the centers of two touching spheres are spaced a sphere diameter apart. Third, the just touching condition is captured by two angle conditions: the angles subtended by the two triplets of points (i-1,i,j) and (i,j,j+1) are both 90°. Finally, space-filling requires that the two relevant interaction centers i and j protrude or poke towards each other. Poking can be identified by choosing a reference point i and measuring the mutual distance to other interaction centers (all other points) and finding the nearest local minimum along the string. Were two interaction centers not poking, it would follow that there ought to be a better pair of interacting centers for pinching two parts of the tube together and filling space. A poking interaction constitutes the determination of five pairwise distances [(i,j), (i,j-1), (i,j+1), (i-1,j), and (i+1,j), ensuring that the (i,j) distance is the smallest among them] and masquerading as an effective pairwise interaction.

The rotation angle of the helix is denoted by a variable t, which takes on integer multiples of the rotation angle $\varepsilon_0$ for a discrete string. To go from one point to the next, one rotates about the helical axis (here chosen in the z-direction) by an angle t = $\varepsilon_0$, while simultaneously translating along the axis by the rise per residue p. In the continuum limit, any generic point of the space-filling helix, here denoted by t=0, has poking (protruding) contacts with a pair of points with t=+ t* or – t* with $t_{deg}^*$=302.3891….° or equivalently $t^*_{rad}$=5.2777…. radians. The tube cross section centered at t=0 (we will call this a coin) just touches each of the two coins centered at t=+ t* and – t*, thereby filling the space between them. And this happens for each of the infinite coins in the continuum limit.

For a fixed bond length b, a helix is completely characterized by the rotational angle $\varepsilon_0$ and the rise per residue p, see Figure SI 1 in Supplementary Information for the definition of these quantities. The pitch of the helix is given by P = p (360°/ $\varepsilon_0$). Defining the helix radius to be R, the coordinates of the points lying along the helical string are, with integer n,

$x_n$ = R cos((n-1) $\varepsilon_0$)
$y_n$ = R sin((n-1) $\varepsilon_0$)                                                                                                       (1)
$z_n$ = P (n-1) $\varepsilon_0$/(2π).

As mentioned before, Δ = $R_{curv}$ = R(1+$\eta^2$), where η=P/(2πR). The space-filling engendered by a pair of poking contacts is enforced by two requirements: the distance between the poking pair (i,i+m) is required to be equal to the coin diameter 2Δ = 2R(1+$\eta^2$), and the lines joining [(i and i+m) and (i+m and i+m+1)] and [(i and i+m) and (i-1 and i)] are perpendicular to each other. For a discrete chain, what we have is a plausible condition for space-filling that has the correct limiting behavior in the continuum [14]. The two conditions are tantamount to the two coins at i and (i+m) barely touching end on end and correspond to the geometrical criterion for determining the closest distance between two skew lines with a need for both distance and angle constraints. For every integer m>2, these two conditions can be satisfied by judiciously choosing the two variables $\varepsilon_0$ and η appropriately. And there is no solution when m=1 or m=2.

Each interaction center i can be imagined to be associated with two coins (segments with a tube symmetry), both of radius $\Delta$, centered around it and indicating the directions of its neighbors along the string. The face of the first coin is thus normal to the (i,i-1) bond direction and that of the second coin is normal to the (i,i+1) bond direction, thereby completely specifying the string conformation.

Defining m to be the sequence separation between i and j, we seek helices with pairs of poking contacts (i,i-m) and (i,i+m) for every i. Here m is an integer, which operationally takes on one of the values 3, 4, 5, … The first two integer values (m=1 and m=2) are excluded because they cannot fill space. We ensure maximal sustained self-interactions of the string by pairing the coin at i pointing to (i-1) to the coin at (i+m) pointing to (i+m+1). The coin at i pointing to (i+1) is likewise paired with the coin at (i-m) pointing to (i-m-1) (Figure 3). Because of helix periodicity, the same arrangement repeats for every i. The successive turns of the helical tube stack atop each other and track each other. Their envelopes smoothly fit alongside and next to each other to fill the space between them and pack efficiently (Figure 4a). Successive turns of the space-filling helix stack atop each other, track each other, and pack efficiently. Furthermore, there is no steric clash or overlaps of any pair of coins in any of the discrete variants or in the continuum limit.

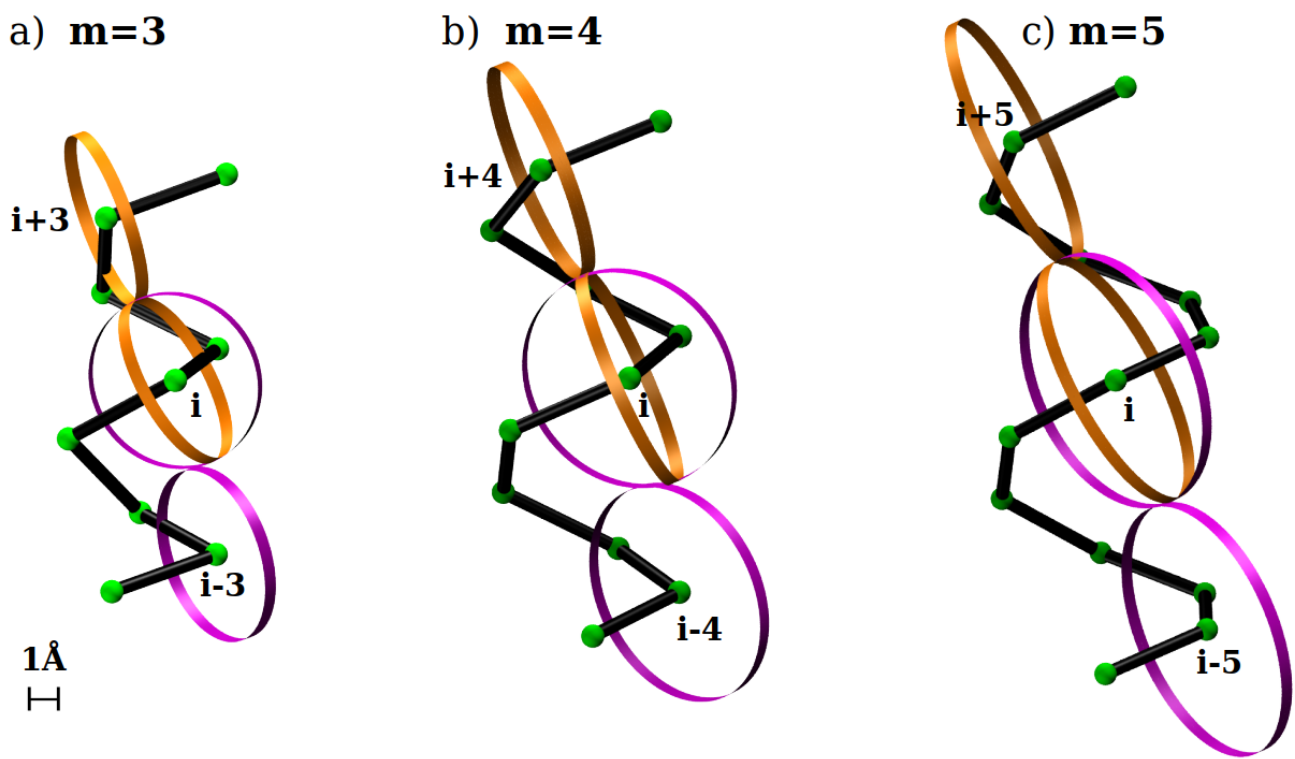

Figure 3: Sketches of discrete space-filling helices, with bond length b = 3.81 Å, corresponding to poking contacts between points i and i+m, with m = 3, 4, and 5. The two coins (in purple and in orange) at point i on the string are shown. a) For clarity, we show just one coin for point i-3 (in purple) and for point i+3 (in orange). The pair of purple coins form a poking contact as do the pair of orange coins. b) and c) show discrete space-filling helices with protruding contacts of the types (i,i+4) (m=4) and (i,i+5) (m=5), respectively. These are the first three space-filling discrete helices in a countable set.

In all three cases, the distance between points i and i+m is 2Δ (the coin diameter), where Δ is the radius of curvature of the helical curve. The line joining the pair of points (i,i+m) is perpendicular to the two straight lines joining the points (i+m,i+m+1) and (i-1,i). These two conditions ensure the maximization of the non-local self-interaction (non-local space-filling condition), while the condition Δ=$R_{curv}$ (thickness equal to the radius of curvature of the helix) maximizes the self-interaction at the local level [14] (local space-filling condition). These local and non-local space-filling conditions cannot be satisfied simultaneously for m=1 and m=2.

The top part of Figure 4 shows side views and the bottom and top views, for each of the discrete helices depicted in Figure 3. Even though all m = 3, 4, 5, … helices are geometrically space-filling, the m=3 helix emerges as the sole contender for the protein α-helix because of the snug fit of the backbone atoms within the tube. The helices, other than m=3, when dressed in protein backbone atoms have a hole in the center (as seen in the top and bottom views of these helices) and empty space in the region between consecutive helical turns (side views). The m=3 helix corresponds to poking contacts between i and (i+3), which is perfectly realized in protein helices and corresponds to the thinnest string (for bond length b=3.81Å) that can form a space-filling helix (the thickness Δ as a function of m is presented in Table 1).

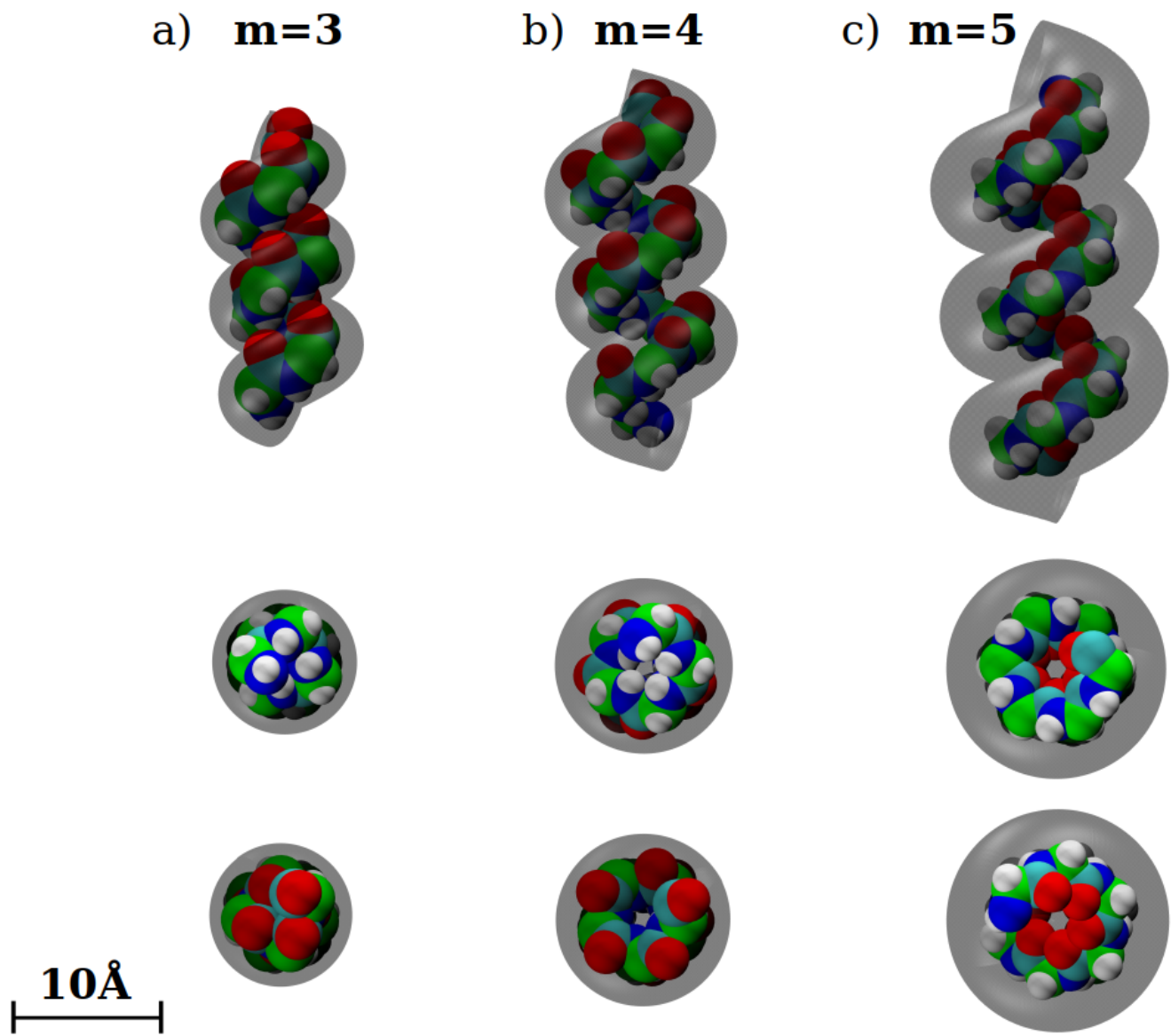

Figure 4: Sketches of the discrete helices shown in Figure 3 dressed in protein backbone atoms employing the PULCHRA software [25]. All spheres have van der Waals radii of the corresponding atom type. Each of the three discrete helices decorated with backbone atoms is shown in the top row in a side view, and in the next two rows in a top and bottom view, respectively. $C_\alpha$ atoms are shown in green, carbon atoms other that $C_\alpha$ are turquoise, oxygen atoms are red, nitrogen atoms are dark blue, while hydrogen atoms are shown in white. The tube envelopes are shown in gray with appropriate radius (equal to the relevant coin thickness $\Delta$ for the respective m-value) presented in Table 1.

As observed in nature, we find that the space-filling discrete helix does not have an integer number of residues per turn. The amino acid side chains stick out in the approximately negative normal direction [26] in a Frenet coordinate system [24] of the backbone and stay out of the way of the backbone atoms and each other but play a vital role in the interactions of a helix with its surroundings. We note that there are just three independent variables that define a helix: the set (b, $\varepsilon_0$, and p) is one employed by Pauling [1]. Other characteristics (see Figure SI 1 for definitions) like the bond bending angle $\theta$ and the dihedral angle $\mu$ characterizing the local chain geometry can be deduced straightforwardly from the triplet of independent variables. The attributes of the m=3 space-filling helix, the Pauling hydrogen-bonded helices [1], and the experimentally determined α-helix are presented in Table 1. The molecular structures of the m=3 space-filling helix, the Pauling α-helix, and the protein helix are all compatible with each other.

| Quantity | Protein helices | Pauling α-helix (m=3) | Pauling γ-helix (m=5) | m = 3 | m = 4 | m = 5 | Asymptotic behavior | Continuum space-filling helix (m = ∞) |
|---|---|---|---|---|---|---|---|---|
| d(i,i+m)/2Δ | 0.97 ± 0.05 | 0.96 | 0.73 | 1 | 1 | 1 | 1 | 1 |
| ∠(i-1,i,i+m) [°] | 86.6 ± 3.5 | 83.2 | 100.0 | 90 | 90 | 90 | 90 | 90 |
| $\varepsilon_0$ [°] | 99.1 ± 3.1 | 97.6 | 70.2 | 99.8 | 76.1 | 61.3 | $t^*_{deg}/m$ | 0 |
| n=360°/$\varepsilon_0$ | 3.63 ± 0.10 | 3.69 | 5.13 | 3.61 | 4.73 | 5.87 | m (360°/$t^*_{deg}$) | ∞ |
| p/b | 0.39 ± 0.02 | 0.39 | 0.26 | 0.41 | 0.40 | 0.39 | $\eta/\sqrt{(1+\eta^2)}$ | 0.37 |
| P/b | 1.43 ± 0.06 | 1.42 | 1.33 | 1.49 | 1.87 | 2.27 | $2\pi\eta m/(t^*_{rad}\sqrt{(1+\eta^2)})$ | ∞ |
| R/b | 0.60 ± 0.02 | 0.61 | 0.84 | 0.60 | 0.75 | 0.90 | $m/(t^*_{rad}\sqrt{(1+\eta^2)})$ | ∞ |
| η | 0.38 ± 0.02 | 0.37 | 0.25 | 0.3997 | 0.3997 | 0.3996 | - | 0.3998 |
| Δ/b | 0.69 ± 0.02 | 0.70 | 0.89 | 0.69 | 0.86 | 1.05 | $m\sqrt{(1+\eta^2)}/t^*_{rad}$ | ∞ |
| **b = 3.81 ± 0.02 Å** | | | | | | | | |
| p [Å] | 1.51 ± 0.07 | 1.47 | 0.99 | 1.58 | 1.51 | 1.47 | | |
| P [Å] | 5.46 ± 0.22 | 5.42 | 5.08 | 5.69 | 7.13 | 8.65 | | |
| R [Å] | 2.30 ± 0.06 | 2.34 | 3.20 | 2.27 | 2.84 | 3.45 | | |
| Δ [Å] | 2.63 ± 0.06 | 2.66 | 3.40 | 2.63 | 3.29 | 4.00 | | |
| θ [°] | 91.3 ± 2.0 | 92.11 | 112.6 | 91.8 | 111.0 | 123.9 | | |
| μ [°] | 49.7 ± 3.3 | 47.54 | 20.7 | 52.4 | 34.4 | 25.8 | | |

Table 1: Comparison of dimensionless quantities (lengths are measured in units of the bond length b) characterizing the geometries of protein helices, the two Pauling helices, the theoretical predictions of space-filling m=3, 4, 5 discrete helices, and in the continuum limit with m approaching ∞. Here

$t^*_{deg}$=302.3891…. °, $t^*_{rad}$=5.2777…. rad (see text). Our space-filling hypothesis is expressed by the conditions given in the first two rows and are used to solve for the characteristics of the geometrically derived helices. The empirical data for proteins in the second column was obtained from an analysis of 39,922 quartets embedded within α-helices in 4391 high-resolution globular protein structures [27]. By embedded, we mean that the 4 $C_\alpha$ atoms within a quartet were enveloped by a full complement of protruding interactions *and* hydrogen bonds. The data for the two Pauling helices were derived from Pauling's original paper [1] with a bond length equal to 3.81 Å and the rise per residue equal to (1.47Å, 0.99Å) and the number of residues per turn (n=360°/$\varepsilon_0$) equal to (3.69, 5.13) for Pauling's α-helix and γ-helix respectively. The theoretically predicted geometries of the space-filling m=3 helix and Pauling's α-helix are compatible with each other and with empirical data, within error bars. The values of η=P/(2πR) of the space-filling discrete helices are very nearly, but not *exactly*, independent of m. The values of the rotation angles are approximately equal to $t^*$/m confirming that the discrete case entails fitting of m coins in the rotation angle interval between 0 and $t^*$. This is analogous to the conditions of quantization in a Bohr atom.

To reiterate, the helix attributes are derived by solving for two variables η=P/(2πR) and $\varepsilon_0$ in the coordinates of the points lying along the helical string (Equation (1)) that satisfy two equations simultaneously. The first sets the angle ∠(i-1,i,i+m) to be 90°. The second sets the distance d(i,i+m) between two protruding points, i and i+m, to be 2R(1+η²), where η=P/(2πR). Operationally, it is simplest to measure all lengths in units of R and then deduce all attributes of the helix from the solution. We ensure that both equations are simultaneously satisfied with a relative error less than 1 part in a million for each condition.

From among the infinite countable number of candidate space-filling helices, just one (m=3) fits the bill for the protein α-helix. This is because the protein backbone atoms approximately fill the space within the tube only for this case. Each of the m>3 cases would still be space-filling unto itself but would have extra room within the helix for additional backbone atoms (Figure 4). A helix is a space-filling string under two conditions: it must have one of the quantized Goldilocks ratios of Δ/b that corresponds to one of the integer values of m (see Table 1); and once the bond length b is specified, the appropriate Goldilocks tube thickness must be Δ. It is remarkable that *both* these 'improbable' conditions hold approximately for proteins corresponding to the m=3 space-filling string.

The space-filling in a protein corresponds to the need for water being squeezed out between successive helical turns to occlude hydrophobic exposed areas. Just as each point on the string has two interacting coins, hydrogen bonding at the backbone level also originates from two distinct points of interaction, a donor (-NH amide group) and an acceptor (-C=O carbonyl group) in the peptide main chain. We have previously analyzed [21,22] 3594 α-helices in 4391 high-resolution protein structures [27] and find that there is a two-way association between hydrogen bonds and protruding interactions in most cases in α-helices. This is in accord with the well accepted notion that while the hydrophobic effect is a driver of protein folding, the internal structure and the rigidity of the helix is provided by hydrogen bonds. The m=3 protruding interaction is commonly alluded to in the literature as an (i,i+4) hydrogen bond.

## m=3 Space-filling helix

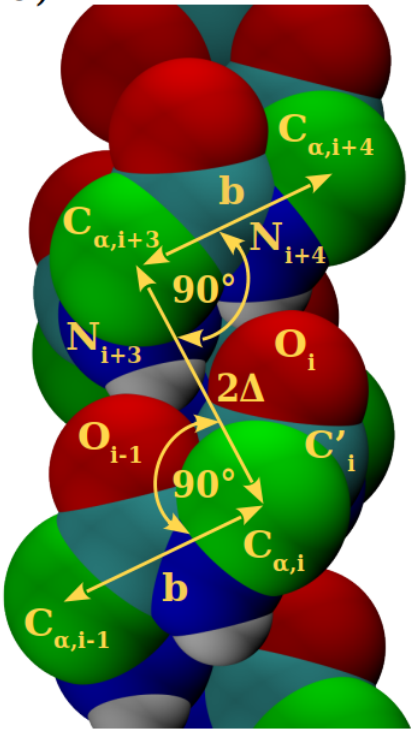
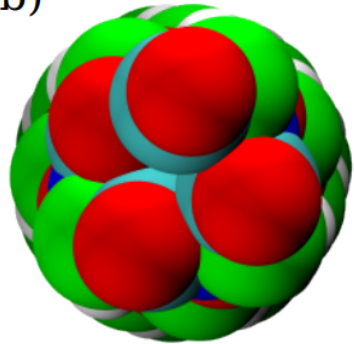
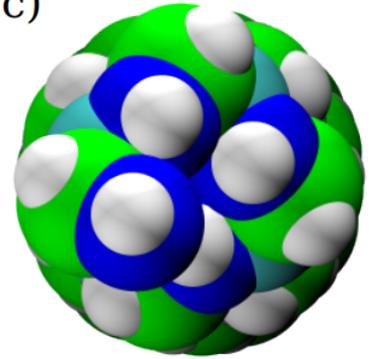

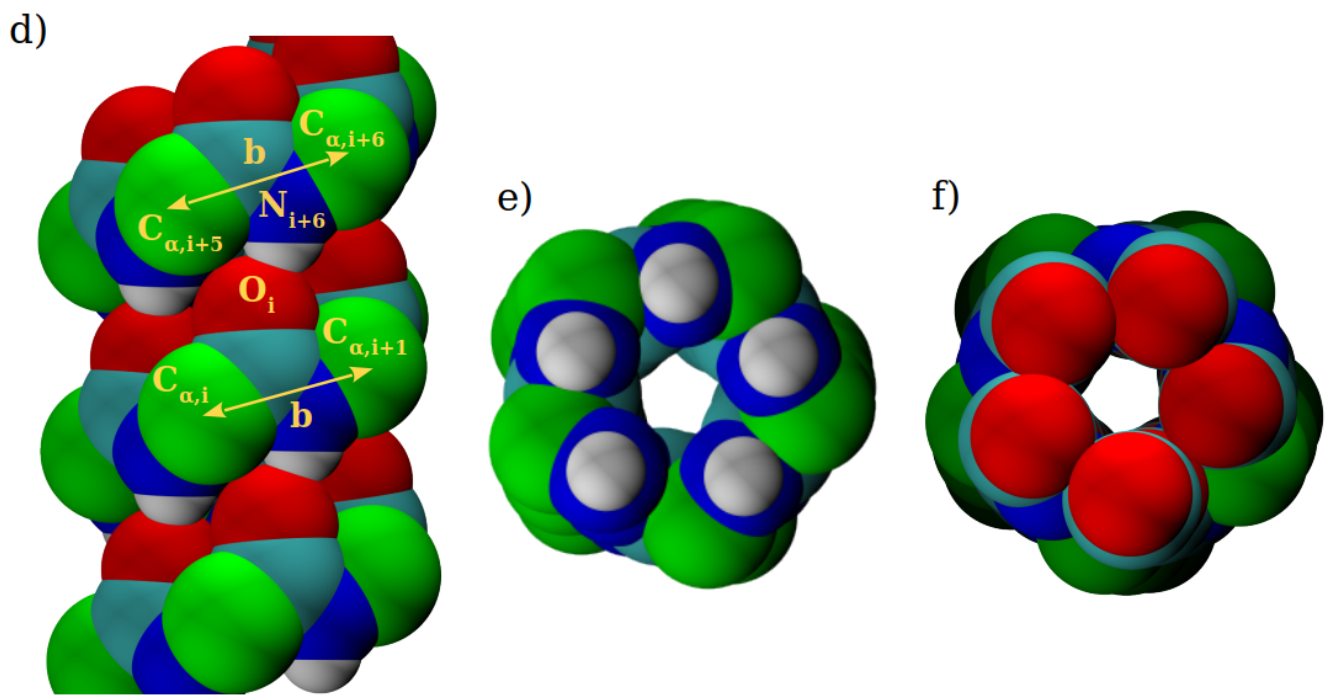

Figure 5: Sketches of the m=3 space-filling helix and the Pauling γ-helix. The molecular structure of the helix is derived at the level of C-alpha atoms and 'decorated' [25] by the inclusion of backbone atoms (see text). There is a protruding contact between i and i+3 in the helix. Hydrogen bonds formed between [$O_i$ and $N_{i+4}$] and [$O_{i-1}$ and $N_{i+3}$] bracket the protruding contact and provide support. The non-bonded backbone atoms are effectively hard spheres. There is space between the van der Waals spheres of the $N_{i+3}$ - H atom group and both the $O_i$ and the $C_i'$ atoms, allowing room for squeezing the helix pitch, as observed in Table 1. The Pauling γ-helix is derived from data presented by Pauling [1] and again decorated by backbone atoms and is *not* space-filling (see text). $C_α$ atoms are shown in green, carbon atoms other that $C_α$ are turquoise, oxygen atoms are red, nitrogen atoms are dark blue, while hydrogen atoms are shown in gray. The white pentagon denotes a hole running through the interior of the helix.

The Pauling γ-helix is not observed in nature despite forming hydrogen bonds in a concerted manner between the backbone atoms. The hydrogen bond was predicted to be between the N-H group of one amino acid and the C=O group of the amino acid *six* residues earlier. Figure 5 illustrates three views (side, top, and bottom) each of the molecular structures of the m=3 space-filling helix and Pauling's γ-helix. There are (i,i+5) protruding contacts in the γ-helix. However, there is significant overlap between the coins at i and i+5 and the (i,i+5,i+6) angle is around 100°, instead of the canonical 90°. The pentagonal shaped hole in the center of the γ-helix underscores the conspicuous absence of space-filling in the γ-helix. Thus, the requirement of filling space would have ruled out Pauling's γ-helix in favor of his α-helix.

The virtually perfect fit of quantum chemistry to geometry is illustrated in Figures 1b and 5a, which shows the molecular geometry of the peptide bond and an α-helix. The geometrical constraints on the space-filling helix is that the (i-1,i) bond of length b is perpendicular to the straight line joining i to i+3, and that the (i,i+3) distance has a length 2Δ roughly equal to 5.26Å (see Table 1), the diameter of the space-filling coin (Figure 2a). Likewise, the planar peptide bond in Figure 1b is a space-filling plane with perpendicular sides b and W. The dependence of the orientation of peptide groups (interpreted as peptide-group dipoles) on the backbone geometry has been found to be important for the development of powerful coarse-grained potentials for understanding protein secondary structure formation [29,30]. The value of W turns out to be around 5.24Å and is roughly equal to 2Δ in excellent accord with the geometrical requirements of space-filling. We predict, from space-filling considerations, that the peptide plane between the (i-1) and i $C_\alpha$ atoms ought to extend in the direction of the (i+3) $C_\alpha$ atom poking with i. We find, strikingly, that the (i+3) $C_\alpha$ atom essentially lies in the peptide plane [i-1,i] with an absolute distance of 0.06 ± 0.04 Å from it.

Protein helices are predominantly right-handed because of the occurrence of left-handed amino acids. This leads to steric clashes of oxygen backbone atoms with the side chain atoms in a left-handed helix [8,31]. Just as in the original Pauling analysis [1], there is no overt chiral symmetry breaking in our theory.

**Perspective**

**Discreteness leads to quantization of helices**

Discreteness is at the heart of quantum mechanics. The energy of an electron in a Bohr hydrogen atom can only take on certain discrete values. This arises in a hand-waving semi-classical description from an integer number of wavelengths of matter waves fitting within the circumference of the orbit of an electron. Here we see a similar situation but in a vastly less profound context. The discrete helices arise by fitting in a sequence separation of integer m between a pair of protruding contacts. The *universal* value of $t^*$=302.3891…. ° plays a central role here.

**Pauling and Kepler**

One of Pauling's two predictions is *the* correct solution for the structure of the α-helix, which satisfies the varied dictates of chemistry. Our theory is instead based on just one hypothesis that the discrete helix is a variant of the space-filling continuum helix, a la Kepler [12]. Yet, the Pauling and Kepler helices are entirely compatible with each other and with data. One might wonder whether this is an example of universality in critical phenomena [32], in which, for example, the critical exponents of the liquid-vapor transition are the same as that of the consolute point of two fluids. *It is not.* Here we are dealing with the structure at the molecular scale and not the long length scale behavior. The Pauling helix and the space-filling helix are two distinct ways of looking at the problem and ought to be compatible with each other provided a protein is subject to both the laws of chemistry and to the geometrical space-filling requirement.

Pauling [1] used "*the complete and accurate determination of the crystal structure of amino acids, peptides, and other simple substances related to proteins …. to construct two reasonable hydrogen-bonded* *helical configurations for the polypeptide chain"*. In contrast, our theory remarkably has no input and no adjustable parameters, other than the space-filling hypothesis for the helix. It is extraordinary that the molecular structure of a space-filling helix along with the string thickness can be predicted exactly without any other chemical input other than the Pauling value of the $C_\alpha$-$C_\alpha$ pseudo-bond length that sets the length scale in the theory (see Table 1).

**The protein free energy landscape**

The free energy landscape [33-35] is a powerful concept that links protein sequence and structure by showing how the energetic landscape guides a protein towards its native, functional conformation. The general belief is that the sequence determines the energy landscape; different sequences will have different landscapes. The native structure is at the bottom of the funnel for that specific sequence with the shape of the landscape (e.g., ruggedness, depth of the funnel) affecting the folding process. Our analysis of the space-filling helix arising from geometry and being independent of sequence suggests a simpler scenario in which the free energy landscape is sculpted in two distinct steps [36]. The first involves the common character of all proteins with the landscape comprising thousands of minima corresponding to all putative native state structures. The putative menu of native state folds is created by considerations of space-filling with poking contacts playing a role when relevant. The second step is the fitting of a specific sequence to one of these pre-sculpted minima yielding a funnel landscape promoting rapid folding. In this way of thinking, the textbook wisdom [28] of sequence determining structure ought to be replaced by sequence selecting the best fit structure. The number of sequences is vastly larger than the number of distinct folds [37,38]. A striking illustration is provided by the globin protein family [39], wherein a highly conserved three-dimensional fold is maintained across members despite substantial divergence at the primary sequence level. This is loosely analogous to a fixed menu of crystalline structures [13] determined from considerations of geometry and symmetry with a specific chemical selecting its ground state from this predetermined menu.

**Conclusions**

Here we have studied just one building block of protein structures, the helix. In addition to all his seminal work developing quantum chemistry, Pauling also worked out the structures of two types of sheet structures comprised of hydrogen-bonded strands [40]. Our success here in predicting the molecular structure of the protein α-helix using just the bond length as the sole input suggests that understanding the lively interplay between geometry and chemistry that rules the behavior of proteins is an interesting challenge that could provide a route to creating a unified framework for understanding proteins.

**Acknowledgements**

We are indebted to George Rose for his collaboration and inspiration.

**Conflicts of Interest Statement**

The authors have no conflicts of interest to declare.

**Funding Statement**

JRB was supported by a Knight Chair at the University of Oregon. AG acknowledges the support from MIUR PRIN-COFIN2022, under Grant No. 2022JWAF7Y. TXH is supported by the Vietnam National Foundation for Science and Technology Development (NAFOSTED), under Grant No. 103.01-2023.110. This project received funding from the European Union's Horizon 2020 research and innovation program under Marie Skłodowska-Curie Grant Agreement No. 894784 (TŠ). The authors acknowledge Research Advanced Computing Services (RACS) at the University of Oregon for providing computing resources that have contributed to the research results reported within this publication.

**Author Contributions**

JRB, AM, TŠ developed the ideas in this paper. TŠ carried out the calculations and prepared the figures with advice from JRB. JRB wrote the paper. All authors read and helped improve the manuscript.

**Data Availability Statement**

The data that support the findings of this study are available upon reasonable request from the corresponding author.

**Supplementary Material**

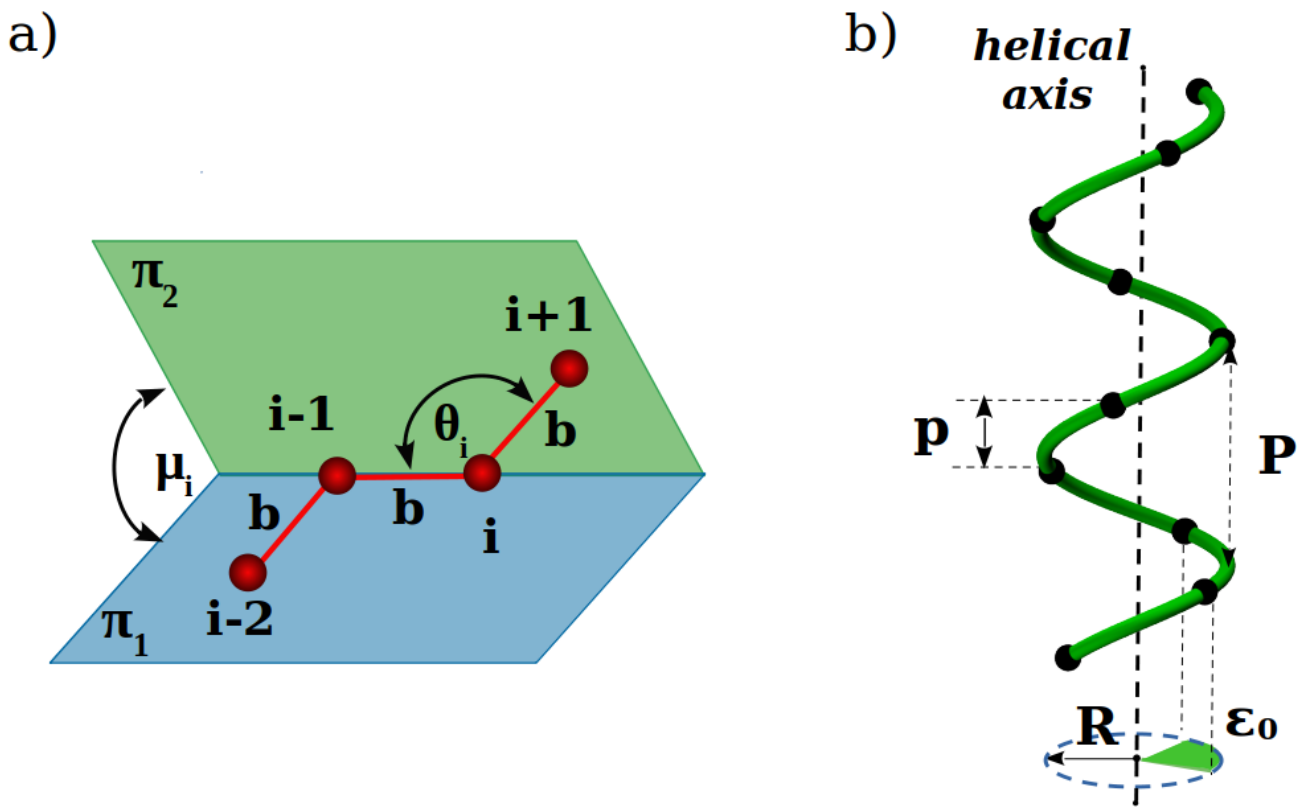

Figure SI 1: Definitions of relevant variables. (a) The bond bending angle $\theta_i$ at position i is the angle subtended at i by points (i-1) and (i+1) along the string. The dihedral angle $\mu_i$ is the angle between the planes $\pi_1$ and $\pi_2$ formed by [(i-2), (i-1), i] and [(i-1),i,(i+1)], respectively. The bond length b is the distance between any two neighboring points along the string. (b) Sketch of a helical curve (in green) with equidistant discrete points (in black) along the curve. The radius R and pitch P of the helix are indicated. For a given discretization of a helical curve, in order to go from one discrete point to the next, one rotates about the helical axis (thick black dashed vertical line) by the characteristic rotation angle denoted by $\varepsilon_0$, while simultaneously translating along the helical axis by the rise per residue denoted by p.